%% file: pol06.tex
\documentclass[vecphys]{svmult}

\usepackage{makeidx}         
\usepackage{graphicx}        
\usepackage{multicol}        

\makeindex             

\begin{document}
\title*{Path Integral Methods in the Su-Schrieffer-Heeger Polaron Problem}
\author{Marco Zoli}
\institute{Istituto Nazionale Fisica della Materia - Dipartimento
di Fisica, Universit\'a di Camerino, 62032 Camerino, Italy.
\texttt{ marco.zoli@unicam.it}}
%
%
\maketitle

{\bf Summary} I propose a path integral description of the
Su-Schrieffer-Heeger Hamiltonian, both in one and two dimensions,
after mapping the real space model onto the time scale. While the
lattice degrees of freedom are classical functions of time and are
integrated out exactly, the electron particle paths are treated
quantum mechanically. The method accounts for the variable range
of the electronic hopping processes. The free energy of the system
and its temperature derivatives are computed by summing at any $T$
over the ensemble of relevant particle paths which mainly
contribute to the total partition function. In the low $T$ regime,
the {\it heat capacity over T} ratio shows un upturn peculiar to a
glass-like behavior. This feature is more sizeable in the square
lattice than in the linear chain as the overall hopping potential
contribution to the total action is larger in higher
dimensionality. The effects of the electron-phonon anharmonic
interactions on the phonon subsystem are studied by the path
integral cumulant expansion method.

\section{Introduction}
\label{sec:1}

There has been a growing interest towards polarons over recent
years also in view of the technological potential of polymers,
organic molecules
\cite{friend,capasso,nasu,dallak,krish,han,brat}, carbon nanotubes
\cite{veri} and high $T_c$ superconductors \cite{alex} in which
polaronic properties, essentially dependent on the type and
strength of the electron-phonon coupling \cite{alekor,aletv,i1},
have been widely discussed.

One dimensional (1D) systems with a half filled band undergo a
structural distortion \cite{peierls} which increases the elastic
energy and opens a gap at the Fermi surface thus lowering the
electronic energy. The competition between lattice and electronic
subsystems stabilizes the 1D structure which accordingly acquires
semiconducting properties whereas the behavior of the 3D system
would be metallic-like.

Conjugated polymers, for example polyacetylene, show anisotropic
electrical and optical properties \cite{lu} due to intrinsic
delocalization of $\pi$ electrons along the chain of CH units. As
the intrachain bonding between adjacent CH monomers is much
stronger than the interchain coupling the lattice is quasi-1D.
Hence, as a result of the Peierls instability, {\it
trans}-polyacetylene shows an alternation of short and long
neighboring carbon bonds, a dimerization, accompanied by a two
fold degenerate ground state energy. Charge injection in polymers
induces a lattice distortion with the associated formation of
localized excitations, polarons and/or charged solitons. While the
latter can exist only in {\it trans}-polyacetylene, the former
solutions are more general since they do not require such
degeneracy.

The Su-Schrieffer-Heeger (SSH) model Hamiltonian \cite{ssh} has
become a successful tool in polymer physics as it hosts the
peculiar ground state excitations of the 1D conjugated structure
and it accounts for a broad range of polymer properties
\cite{macdiarmid,cat,ssh1}. In the weak coupling regime a
continuum version \cite{tlm} of the SSH model has been developed
and polaronic solutions have been obtained analytically
\cite{campbell}. Although the 1D properties of the SSH model have
been mainly investigated so far
\cite{schulz,fradkin,stafstrom,raedt,zheng,pucci,capone,i2},
extensions to two dimensions were considered in the late eighties
in connection with the Fermi surface nesting effect on quasi 2D
high $T_c$ superconductors \cite{tang}. Recent numerical analysis
\cite{ono} have revealed the rich physics of 2D-SSH polarons whose
mass seems to be larger than in the 1D case, at least in the
intermediate regime of the adiabatic parameter \cite{i3}.

As a fundamental feature of the SSH Hamiltonian the electronic
hopping integral linearly depends on the relative displacement
between adjacent atomic sites thus leading to a nonlocal {\it
e-ph} coupling \cite{sto,per} with vertex function depending both
on the electronic and the phononic wave vector. The latter
property induces, in the Matsubara formalism \cite{mahan}, an
electron hopping associated with a time dependent lattice
displacement. As a consequence time retarded electron-phonon
interactions arise in the system yielding a source current which
depends both on time and on the electron path coordinates. This
causes large {\it e-ph} anharmonicities in the equilibrium
thermodynamics of the SSH model \cite{i4}. Hopping of electrons
from site to site accompanied by a coupling to the lattice
vibration modes is a fundamental process determining the transport
\cite{lang} and equilibrium properties of many body systems
\cite{nieu}. A variable range hopping may introduce some degree of
disorder thus affecting the charge mobility \cite{fish,plyuk} and
the thermodynamic functions.

These issues are hereafter analyzed by means of the path integral
formalism \cite{feynman,devreese,farias,ganbold,raedt1,korn} which
fully accounts for the time retarded {\it e-ph} interactions as a
retarded potential naturally emerges in the exact integral action
\cite{kleinert}. Using the main property of the {\it e-ph}
coupling model, the Hamiltonian linearly dependent on the phonon
displacement field, I attack the SSH model after introducing a
generalized version of the semiclassical model which treats only
the electrons quantum mechanically. Being valid for any {\it e-ph}
coupling value, the path integral method allows one to derive the
partition function and the related temperature derivatives without
those limitations which affect the perturbative studies. The
general formalism, both for the 1D and 2D system, is described in
Section 2 while the path integral approach used to derive the full
partition function of the interacting system is developed in
Section 3. In Section 4, I outline the main features of the
computational method and derive the thermodynamical properties for
a particle described by a SSH Hamiltonian in a bath of harmonic
oscillators. In Section 5, the model is generalized in order to
include the electron-phonon effects on the oscillator bath: the
anharmonic corrections to the total heat capacity are evaluated by
a path integral cumulant expansion in terms of the source current
of the Hamiltonian model. Some conclusions are drawn in Section 6.

\section{The Hamiltonian Model}
\label{sec:2}

In a square lattice with isotropic nearest neighbors hopping
integral $J$, the SSH Hamiltonian for electrons plus {\it e-ph}
interactions reads:

\begin{eqnarray}
H=\,& & \sum_{r,s}\Bigl[(J_{r,s})_x \bigl(f^{\dag}_{r+1,s} f_{r,s}
+ h.c.\bigr)  \, \nonumber \\ &+& (J_{r,s})_y
\bigl(f^{\dag}_{r,s+1} f_{r,s} + h.c.\bigr) \Bigr] \, \nonumber \\
(J_{r,s})_x=\,& & - {1 \over 2}\bigl[ J + \alpha \Delta u_x \bigr]
\, \nonumber
\\ (J_{r,s})_y=\,& & - {1 \over 2}\bigl[ J + \alpha \Delta u_y \bigr]
\, \nonumber \\ \Delta u_x=\, & & u_x(r+1,s) - {u_{x}(r,s)} \,
\nonumber \\    \Delta u_y=\, & & u_y(r,s+1) - {u_{y}(r,s)} \quad,
\, \nonumber \\ \label{eq:1}
\end{eqnarray}

where $\alpha$ is the electron-phonon coupling, ${\bf u}(r,s)$ is
the dimerization coordinate indicating the displacement of the
monomer group on the $(r,s)-$ lattice site, $f^{\dag}_{r,s}$ and
$f_{r,s}$ create and destroy electrons (i.e., $\pi$ band electrons
in polyacetylene). The phonon Hamiltonian is given by a set of 2D
classical harmonic oscillators. The two addenda in
Eq.~(\ref{eq:1}) deal with one dimensional {\it e-ph} couplings
along the {\it x} and {\it y} axis respectively, with first
neighbors electron hopping. Second neighbors hopping processes
(with overlap integral $J^{(2)}$) may be accounted for by adding
to the Hamiltonian the term $H^{(2)}$ such that

\begin{eqnarray}
& &H^{(2)}=\,(J_{r,s})_{x,y} \bigl(f^{\dag}_{r+1,s+1} f_{r,s} +
h.c.\bigr) \, \nonumber \\ & &(J_{r,s})_{x,y}=\,- {1 \over
2}\Bigl[ J^{(2)} + \alpha \sqrt{ (\Delta u_x)^2 + (\Delta u_y)^2 }
\Bigr]\quad. \, \nonumber \\ \label{eq:2}
\end{eqnarray}

The 1D SSH Hamiltonian is obtained by Eq.~(\ref{eq:1}) just
dropping the second term depending on the $y$ coordinate. The real
space Hamiltonian in Eq.~(\ref{eq:1}) can be transformed into a
time dependent Hamiltonian \cite{hamann} by introducing the
electron coordinates: i) $\bigl( x(\tau'),y(\tau') \bigr)$ at the
$(r,s)$ lattice site, ii) $\bigl( x(\tau),y(\tau') \bigr)$ at the
$(r+1,s)$ lattice site and iii) $\bigl( x(\tau'),y(\tau) \bigr)$
at the $(r,s+1)$ lattice site, respectively. $\tau$ and $\tau'$
vary on the scale of the inverse temperature $\beta$. The spatial
{\it e-ph} correlations contained in Eq.~(\ref{eq:1}) are mapped
onto the time axis by changing: $ u_{x,(y)}(r,s) \to
u_{x,(y)}(\tau')$, $u_{x}(r+1,s) \to u_x(\tau)$ and $u_{y}(r,s+1)
\to u_y(\tau)$. Now we set $\tau'=\,0$, $\bigl( x(\tau'),y(\tau')
\bigr) \equiv (0,0)$, $\bigl( u_x(\tau'),u_y(\tau') \bigr) \equiv
(0,0)$. Accordingly, Eq.~(\ref{eq:1}) transforms into the time
dependent Hamiltonian:

\begin{eqnarray}
H(\tau)&=&\,J_x(\tau) \Bigl(f^{\dag}(x(\tau),0)f(0,0) + h.c.
\Bigr)\, \nonumber
\\ &+& J_y(\tau) \Bigl(f^{\dag}(0,y(\tau))f(0,0) + h.c. \Bigr)
\, \nonumber
\\
J_{x}(\tau)&=&\,- {1 \over 2}\bigl[J + \alpha u_{x}(\tau) \bigr]
\, \nonumber
\\
J_{y}(\tau)&=&\,- {1 \over 2}\bigl[J + \alpha u_{y}(\tau)
\bigr]\quad. \, \nonumber
\\ \label{eq:3}
\end{eqnarray}

While the ground state of the 1D SSH Hamiltonian is twofold
degenerate, the degree of phase degeneracy is believed to be much
higher in 2D \cite{ono1} as many lattice distortion modes
contribute to open the gap at the Fermi surface. However, as in
1D, these phases are connected by localized and nonlinear
excitations, the soliton solutions. Thus, also in 2D both electron
hopping between solitons \cite{kivel} and thermal excitation of
electrons to band states may take place within the model. These
features are accounted for by the time dependent version of the
Hamiltonian. As $\tau$ varies continuously on the $\beta$ scale
and the $\tau$-dependent displacement fields are continuous
variables (whose amplitudes are in principle unbound in the path
integral), long range hopping processes are automatically included
in $H(\tau)$ which is therefore more general than the real space
SSH Hamiltonian in Eq.~(\ref{eq:1}) (and Eq.~(\ref{eq:2})). Thus,
by means of the path integral formalism we look at the low
temperature thermodynamical behavior both in 1D and 2D searching
for those features which may be ascribable to some local disorder
related to the variable range of the hopping processes.

The semiclassical nature of the model is evident from
Eq.~(\ref{eq:3}) in which quantum mechanical degrees of freedom
interact with the classical variables $u_{x (y)}(\tau)$. Averaging
the electron operators over the ground state we obtain the time
dependent semiclassical energy per lattice site $N$ which is
linear in the atomic displacements:

\begin{eqnarray}
& &{{<H(\tau)>} \over N}=\,J_{x}(\tau)P\bigl(J, \tau,
x(\tau)\bigr)  + J_{y}(\tau) P\bigl(J, \tau, y(\tau)\bigr) \,
\nonumber \\ & &P\bigl(J, \tau, {\bf v}(\tau)\bigr)=\, {1 \over
{\pi^2}}\int d{\bf k} \cos[{\bf k \cdot v}(\tau)]
\cosh(\epsilon_{\bf k} \tau) n_F(\epsilon_{\bf k}) \quad, \,
\nonumber
\\ \label{eq:4}
\end{eqnarray}

with ${\bf v}(\tau)=\,(x(\tau),0)$ and ${\bf
v}(\tau)=\,(0,y(\tau))$ in the first and second term respectively.
$\epsilon_{\bf k}=\, -J \sum_{i=x,y} \cos(k_i)$ is the electron
dispersion relation and $n_F$ is the Fermi function. The chemical
potential has been pinned to the zero energy level.
Eq.~(\ref{eq:4}) can be rewritten in a way suitable to the path
integral approach by defining

\begin{eqnarray}
& &{{<H(\tau)>} \over N}=\, V\bigl({x}(\tau)\bigr) +
V\bigl({y}(\tau)\bigr) + {\bf u}(\tau) \cdot {\bf j}({\bf
v}(\tau)) \, \nonumber
\\ & &V\bigl({x}(\tau)\bigr)=\,-J P\bigl(J, \tau, {x}(\tau)\bigr)
\, \nonumber
\\ & &V\bigl({y}(\tau)\bigr)=\,-J P\bigl(J, \tau, {y}(\tau)\bigr)
\, \nonumber
\\& &{\bf j}({\bf v}(\tau))= -\alpha P\bigl(J, \tau,
{\bf v}(\tau)\bigr) \, \nonumber \\ & &{\bf
u}(\tau)=\,\bigl(u_{x}(\tau),u_{y}(\tau)\bigr)\quad. \, \nonumber
\\ \label{eq:5}\end{eqnarray}

$V\bigl({x}(\tau)\bigr)$ and $V\bigl(y(\tau)\bigr)$ are the
effective terms accounting for the $\tau$ dependent electronic
hopping while ${\bf j}({\bf v}(\tau))$ is interpreted as the
external source \cite{kleinert} current for the oscillator field
${\bf u}(\tau)$. Averaging the electrons over the ground state we
neglect the fermion-fermion correlations \cite{hirsch} which lead
to effective polaron-polaron interactions in non perturbative
analysis of the model \cite{acqua}. This approximation however is
not expected to affect substantially the following calculations.

\section{The Path Integral Formalism}
\label{sec:3}

Taking a bath of $\bar N$ 2D oscillators, we write the SSH
electron path integral, ${\bf \zeta}(\tau) \equiv
\,\bigl({x}(\tau),{y}(\tau)\bigr)$,  as:

\begin{eqnarray}
<{\bf \zeta}(\beta)&|&{\bf \zeta}(0)>=\,\prod_{i=1}^{\bar N} \int
D{\bf u}_i(\tau)  \int D{\bf \zeta}(\tau)  \, \nonumber
\\
&\cdot& exp\Biggl[- \int_0^{\beta}d\tau \sum_{i=1}^{\bar N} {M
\over 2} \biggl( \dot{\bf u}_i^2(\tau) + \omega_i^2 {\bf
u}_i^2(\tau) \biggr)\Biggr] \, \nonumber
\\
&\cdot& exp\Biggl[- \int_0^{\beta}d\tau \biggl({m \over 2}
\dot{\bf \zeta}^2(\tau) + V\bigl({x}(\tau)\bigr) +
V\bigl({y}(\tau)\bigr) \biggr)\, \nonumber
\\ &+& \sum_{i=1}^{\bar N} {\bf
u}_i(\tau) \cdot {\bf j}({\bf v}(\tau)) \Biggr] \quad,\, \nonumber
\\ \label{eq:6}
\end{eqnarray}

where $m$ is the electron mass, $M$ is the atomic mass and
$\omega_i$ is the oscillator frequency.  As a main feature we
notice that the interacting energy is linear in the atomic
displacement field. Then, the electronic path integral can be
derived after integrating out the oscillator degrees of freedom
which are decoupled along the $x$ and $y$ axis. Accordingly, we
get:

\begin{eqnarray}
& &<{\bf \zeta}(\beta)|{\bf \zeta}(0)>=\,\prod_{i=1}^{\bar N} Z_i
\Biggl[ \int D{x}(\tau) \, \nonumber
\\ &\cdot& exp\Bigl[- \int_0^{\beta}d\tau \biggl({m \over 2}
\dot{x}^2(\tau) + V\bigl({x}(\tau)\bigr) \biggr) - {1 \over \hbar}
A({x}(\tau)) \Bigr] \Biggr]^2 , \, \nonumber \\ & &A({
x}(\tau))=\, -{{\hbar^2} \over {4M}}\sum_{i=1}^{\bar N} {1 \over
{\hbar \omega_i \sinh(\hbar\omega_i\beta/2)}} \, \nonumber \\
&\cdot& \int_0^{\beta} d\tau j({x}(\tau)) \int_0^{\beta}d{\tau''}
\cosh\Bigl(\hbar\omega_i \bigl( |\tau - {\tau''}| - \beta/2 \bigr)
\Bigr) j({x}({\tau''})), \, \nonumber
\\ & &Z_i=\, \Bigl[ {1 \over {2\sinh(\hbar\omega_i\beta/2)}}
\Bigr]^2 \quad.\, \nonumber
\\ \label{eq:7}
\end{eqnarray}

Thus, the 2D electron path integral is obtained after squaring the
sum over one dimensional electron paths.  This permits to reduce
the computational problem which is nonetheless highly time
consuming particularly in the low temperature limit. Note in fact
that the source current $j({x}(\tau))$ requires integration over
the 2D Brillouin Zone (BZ) according to
Eqs.~(\ref{eq:4}),~(\ref{eq:5}) and this occurs for any choice of
the electron path coordinates. The quadratic (in the coupling
$\alpha$) source action $A({x}(\tau))$ is time retarded as the
particle moving through the lattice drags the excitations of the
oscillator fields which take a time to readjust to the electron
motion. When the interaction is sufficiently strong the conditions
for polaron formation may be fulfilled in the system according to
the degree of adiabaticity \cite{brown}. However the present path
integral description is valid independently of the existence of
polarons as it applies also to the weak coupling regime. Assuming
periodic conditions ${x}(\tau)=\, {x}(\tau + \beta)$, the particle
paths can be expanded in Fourier components

\begin{eqnarray}
& &{x}(\tau)=\,{x}_o + \sum_{n=1}^\infty 2\Bigl(\Re {x}_n \cos(
\omega_n \tau) - \Im {x}_n \sin( \omega_n \tau) \Bigr)\, \nonumber
\\ & &\omega_n=\,2\pi n/\beta \label{eq:8}
\end{eqnarray}

and the open ends integral over the paths $\int D{x}(\tau)$
transforms into the measure of integration $\oint D{x}(\tau)$.
Taking:

\begin{equation}
\oint D{x}(\tau)\equiv \int_{-\infty}^{\infty}{{d{x}_o} \over
{\bigl( 2\pi\hbar^2/mK_BT \bigr)^{(1/2)}}} \prod_{n=1}^{\infty}
\Biggl[{{\int_{-\infty}^{\infty} d\Re {x}_n
\int_{-\infty}^{\infty} d\Im {x}_n} \over {\bigl( \pi \hbar^2
K_BT/m\omega_n^2 \bigr)}} \Biggr] \quad,\label{eq:9}
\end{equation}

we proceed to integrate Eq.~(\ref{eq:7}) in order to derive the
full partition function of the system versus temperature.

Let's point out that, by mapping the electronic hopping motion
onto the time scale, a continuum version of the interacting
Hamiltonian (Eq.~(\ref{eq:3})) has been de facto introduced.
Unlike previous \cite{lu} approaches however, our path integral
method is not constrained to the weak {\it e-ph} coupling regime
and it can be applied to any range of physical parameters.

\section{Computational Method and Thermodynamical Results}
\label{sec:4}

As a preliminar step we determine, for a given path and at a given
temperature: i) the minimum number ($N_{\bf k}$) of ${\bf
k}-$points in the BZ to accurately estimate the average
interacting energy per lattice site and, ii) the minimum number
($N_{\tau}$) of points in the double time integration to get a
numerically stable source action in Eq.~(\ref{eq:7}). The momentum
integrations required by Eq.~(\ref{eq:4}) converge by summing over
1600 and 70 points in the reduced 2D and 1D BZ, respectively.
Moreover, $N_{\tau}=\,300$ at $T=1K$.

Computation of Eqs.~(\ref{eq:4}) - ~(\ref{eq:7}) requires fixing
two sets of input parameters. The first set contains the physical
quantities characterizing the system: the bare hopping integral
$J$, the oscillator frequencies $\omega_i$ and the effective
coupling $\chi=\, \alpha^2 \hbar^2 /M$ (in units $meV^3$). The
second set defines the paths for the particle motion which mainly
contribute to the partition function through: the number of pairs
($\Re x_n, \Im x_n$) in the Fourier expansion of Eq.~(\ref{eq:8}),
the cutoff ($\Lambda$) on the integration range of the expansion
coefficients in Eq.~(\ref{eq:9}) and the related number of points
($N_\Lambda$) in the measure of integration which ensures
numerical convergence.

After introducing a dimensionless path

\begin{equation}
x(\tau)/a=\,{\bar x_o} + \sum_{n=1}^{N_p} \Bigl(\bar a_n \cos(
\omega_n \tau) + \bar b_n \sin( \omega_n \tau) \Bigr) \quad,
\label{eq:10}
\end{equation}

with: $\bar x_o \equiv x_o /a$, $\bar a_n \equiv 2 \Re x_n /a$ and
$\bar b_n \equiv -2 \Im x_n /a$, the functional measure of
Eq.~(\ref{eq:9}) can be rewritten for computational purposes as:

\begin{eqnarray}
\oint Dx& &(\tau) \approx {a \over {\sqrt{2}}}\biggl({a \over
2}\biggr)^{2N_p} {{(2\pi \cdot 4\pi \cdot \cdot \cdot 2N_p \pi)^2}
\over { (\pi\hbar^2/mK_BT)^{N_p + 1/2}}} \int_{-\Lambda
/a}^{\Lambda /a}{d \bar x_o} \cdot \, \nonumber \\ &
&\int_{-2\Lambda /a}^{2\Lambda /a}{d \bar a_1} \int_{-2\Lambda
/a}^{2\Lambda /a}{d \bar b_1} \cdot \cdot \cdot \int_{-2\Lambda
/a}^{2\Lambda /a}{d \bar a_{N_p}} \int_{-2\Lambda /a}^{2\Lambda
/a}{d \bar b_{N_p}} \quad.\, \nonumber \\ \label{eq:11}
\end{eqnarray}

In the following we take the lattice constant $a=\,1 \AA$. As a
criterion to set the cutoff $\Lambda$ on the integration range, we
notice that the functional measure normalizes the kinetic term in
Eq.~(\ref{eq:7}):

\begin{equation}
\oint Dx(\tau) exp\Bigl[-{m \over 2}\int_0^\beta d\tau
\dot{x}^2(\tau)\Bigr]\equiv 1  \label{eq:12}
\end{equation}

and this condition holds for any number of pairs $N_p$ truncating
the Fourier expansion in Eq.~(\ref{eq:9}). Then, taking $N_p=\,1$,
the left hand side of Eq.~(\ref{eq:12}) transforms as:

\begin{eqnarray}
& &\oint Dx(\tau) exp\Bigl[-{m \over 2}\int_0^\beta d\tau
\dot{x}^2(\tau)\Bigr] \simeq {4 \over \pi} \Bigl[\int_{0}^U dy
exp(-y^2)\Bigr]^2 \, \nonumber \\ & &U \equiv \sqrt{2
\pi^3}{\Lambda \over \lambda} \, \nonumber \\ & & \lambda=\,
\sqrt{{2 \pi \hbar^2} \over {m K_BT}} \quad. \label{eq:13}
\end{eqnarray}

Using the series representation \cite{grad}

\begin{equation}
\int_{0}^U dy exp(-y^2)=\,\sum_{k=0}^\infty {{(-1)^k U^{2k+1}}
\over {k! (2k+1)}} \quad,\label{eq:14}
\end{equation}

one determines $U$ (after setting the series cutoff $k_{max}$
which ensures convergence) by fitting the Poisson integral value
$\sqrt{\pi}/2$.

Thus, we find that the cutoff $\Lambda$ can be expressed in terms
of the thermal wavelength $\lambda$ as $\Lambda \sim  \lambda /
\sqrt{2 \pi^3}$ hence, it scales versus temperature as $\Lambda
\propto 1/\sqrt{T}$. This means physically that, at low
temperatures, $\Lambda$ is large since many paths are required to
yield the correct normalization. For example, at $T=\,1K$, we get
$\Lambda \sim 284 \AA$. Numerical investigation of
Eq.~(\ref{eq:7}) shows however that a much shorter cutoff suffices
to guarantee convergence in the path integral, while the cutoff
temperature dependence implied by Eq.~(\ref{eq:12}) holds also in
the computation of the interacting partition function. The
thermodynamical results hereafter presented have been obtained by
taking $\Lambda \sim \lambda /(10 \sqrt{2 \pi^3})$. Summing in the
1D system over $N_\Lambda \sim 20 /\sqrt{T}$ points for each
integration range and taking $N_p=\,2$, we are then evaluating the
contribution of $(N_\Lambda + 1)^{2N_p + 1}$ paths (the integer
part of $N_\Lambda$ is obviously selected at any temperature).
Thus, at $T=\,1K$ and in 1D, we are considering $\sim 4 \cdot
10^6$ different paths for the particle motion while, at $T=\,100K$
the number of paths drops to 243 \cite{i5}. In the 2D problem,
$N_\Lambda \sim 35 /\sqrt{T}$ points for each Fourier coefficient
are required \cite{i6}. Computation of the second order
derivatives of the free energy in the range $T\in [1,301]K$, with
a spacing of 3K, takes 55hours and 15 minutes on a Pentium 4. Note
that larger $N_p$ in the path Fourier expansion would further
increase the computing time without introducing any substantial
improvement in the thermodynamical output of our calculation.

Although the history of the SSH model is mainly related to wide
band polymers, we take here a narrow band system ($J=\,100meV$)
\cite{i2} with the caveat that electron-electron correlations may
become relevant in narrow bands.

Free energy and heat capacity have been first computed up to room
temperature, both in 1D and 2D, assuming a bath of $\bar N=\,10$
low energy oscillators separated by $2meV$: $\hbar
\omega_1=\,2meV,..., \hbar \omega_{10}=\,20meV$. The lowest energy
oscillator yields the largest contribution to the phonon partition
function mainly in the low temperature regime while the
$\omega_{10}$ oscillator essentially sets the phonon energy scale
which determines the size of the {\it e-ph} coupling. A larger
number $\bar N$ of oscillators in the aforegiven range would not
significantly modify the calculation whereas lower $\omega_i$
values would yield a larger contribution to the phonon partition
function mainly at low $T$.

In the discrete SSH model, the value $\bar \alpha \equiv \,
4\alpha^2/(\pi \kappa J) \sim 1$, marks the crossover between weak
and strong {\it e-ph} coupling, with $\kappa$ being the effective
spring constant. In our continuum and semiclassical model the
effective coupling is $ \chi$. Although in principle, discrete and
continuum models may feature non coincident crossover parameters,
we assume that the relation between $\alpha$ and $J$ obtained by
the discrete model crossover condition still holds in our model.
Hence, at the crossover we get: $\chi_c \sim \, \pi J \hbar^2
\omega^2_{10}/64$. This means that, in Figures 1-3, the crossover
value is set at $\chi_c \sim \,2000meV^3$.

In Fig.1, a comparison between the 1D and the 2D free energies is
presented for two values of $\chi$, one lying in the weak and one
in the strong {\it e-ph} coupling regime. The oscillator free
energies $F_{ph}$ are plotted separately while the free energies
arising from the total action in Eq.~(\ref{eq:7}), shortly termed
$F_{source}$, results from the competition between the free path
action (kinetic term plus hopping potential in the exponential
integrand) and the source action depending on the {\it e-ph}
coupling. While the former enhances the free energy, the latter
becomes dominant at increasing temperatures thus reducing the
total free energy. In general, the 2D free energies have a larger
gradient (versus temperature) than the corresponding 1D terms. The
$F_{ph}$ lie above the $F_{source}$ both in 1D and 2D because of
the choice of the $\hbar \omega_{i}$: in general one may expect a
crossing point between $F_{ph}$ and $F_{source}$ with temperature
location depending on the value of $\chi$. Note that the
$F_{source}$ plots have a positive temperature derivative in the
low temperature regime and this feature is more pronounced in 2D.
In fact, at low $T$, the source action is dominated by the hopping
potential $V\bigl(x(\tau)\bigr)$ while, at increasing $T$, the
{\it e-ph} effects become progressively more important (as the
bifurcation between the $\chi_1$ and $\chi_2$ curves shows) and
the $F_{source}$ get a negative derivative. In 2D, the weight of
the $V\bigl(x(\tau)\bigr)$ term is larger because there is a
higher hopping probability. This physical expectation is taken
into account by the path integral method. At any temperature, we
monitor the ensamble of relevant particle paths over which the
hopping potential is evaluated. For a selected set of Fourier
components in Eq.~(\ref{eq:10}) the hopping decreases by lowering
$T$ but its value is still significant at low $T$. Considered
that: i) the total action is obtained after a $d\tau$ integration
of $V\bigl(x(\tau)\bigr)$ and ii) the $d\tau$ integration range is
larger at lower temperatures, we explain why the overall hopping
potential contribution to the total action is responsible for the
anomalous free energy behavior at low T.

In Fig.2, the heat capacity contributions  due to the oscillators
($C_{ph}$) and electrons plus {\it e-ph} coupling ($C_{sou}$) are
reported on. The values are normalized over $\bar N$. the
previously described summation over a large number of paths turns
out to be essential to recover the correct thermodynamical
behavior in the zero temperature limit. The dimensionality effects
are seen to be large and, for a given dimensionality, the role of
the {\it e-ph} interactions is magnified at increasing $T$. The
total heat capacity ($C_{ph}$ + $C_{sou}$) {\it over} $T$ ratios
are plotted in Fig.3 in the low $T$ regime to emphasize the
presence of an anomalous upturn which appears at $T \preceq 10K$
in 1D and $T \preceq 20K$ in 2D. This feature in the heat capacity
linear coefficient is ultimately related to the sizable effective
hopping integral term $V\bigl(x(\tau)\bigr)$. The strength of the
{\it e-ph} coupling has a minor role in the low $T$ limit although
it determines the shape of the anomaly versus $T$.

Let's focus now on the 1D system and consider the effect of the
oscillators bath on the thermodynamical properties: in Figures 4-6
the ten phonon energies are: $\hbar \omega_1=\,22meV,..., \hbar
\omega_{10}=\,40meV$. Accordingly the crossover is set at $\chi_c
\sim \,8000meV^3$ and three plots out of five lie in the strong
{\it e-ph} coupling regime. As shown in Fig.4, large $\chi$ values
are required to get strongly decreasing free energies versus
temperature while the $\chi=\,3000meV^3$ curve now hardly
intersects the phonon free energy at room temperature. Fig.5 shows
the rapid growth of the source heat capacity versus temperature at
strong couplings whereas the presence of the low $T$ upturn in the
{\it total heat capacity over T} ratio is confirmed in Fig.6. Note
that, due to the enhanced oscillators energies, the phonon heat
capacity saturates here at $T \sim 400K$ (in Fig.2, for the 1D
case, $T \sim 200K$).
\begin{figure}
\centering
\includegraphics[height=9cm,angle=-90]{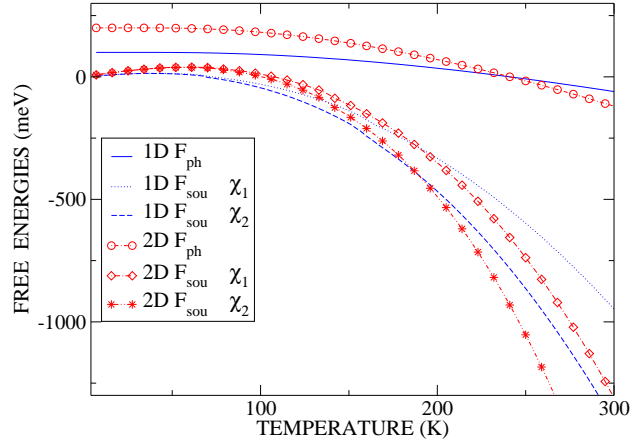}
%
%
\caption{Phonon  ($F_{ph}$) and Source Term ($F_{sou}$)
contributions to the 1D and 2D free energies for two values of the
effective coupling $\chi$: $\chi_1=\,1440 meV^3$ (weak {\it e-ph}
coupling) $\chi_2=\,2560 meV^3$ (strong {\it e-ph} coupling). A
bath of ten phonon oscillators is considered, the largest phonon
energy being $\hbar \omega_{10}=\,20meV$. }
\label{fig:1}       
\end{figure}
\begin{figure}
\centering
\includegraphics[height=9cm,angle=-90]{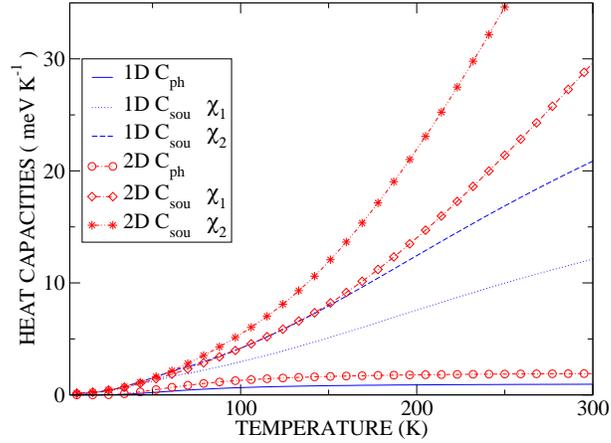}
\caption{Phonon and Source Term contributions (normalized over the
number of oscillators) to the 1D and 2D heat capacities for the
same parameters as in Fig.1. The oscillator heat capacities are
also plotted.}
\label{fig:2}       
\end{figure}
\begin{figure}
\centering
\includegraphics[height=9cm,angle=-90]{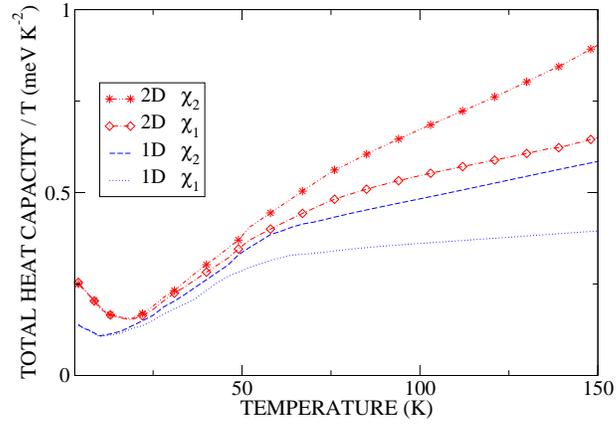}
\caption{Total heat capacity over temperature for the same
parameters as in Fig.1. }
\label{fig:3}       
\end{figure}
\begin{figure}
\centering
\includegraphics[height=9cm,angle=90]{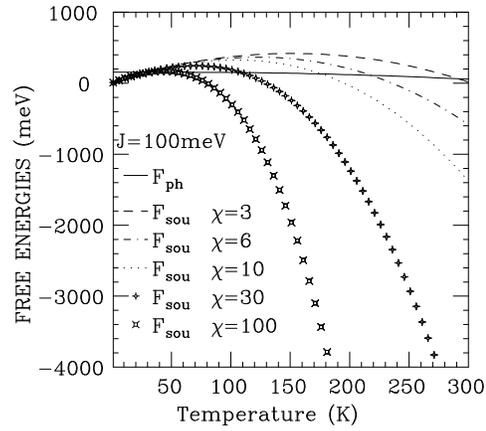}
\caption{1D Phonon and Source Term contributions to the free
energy for five values of the effective coupling $\chi$ (in units
$10^3 meV^3$) and a narrow electron band. A bath of ten phonon
oscillators has been taken, the largest phonon energy is $\hbar
\omega_{10}=\,40meV$.}
\label{fig:4}       
\end{figure}
\begin{figure}
\centering
\includegraphics[height=10cm,angle=90]{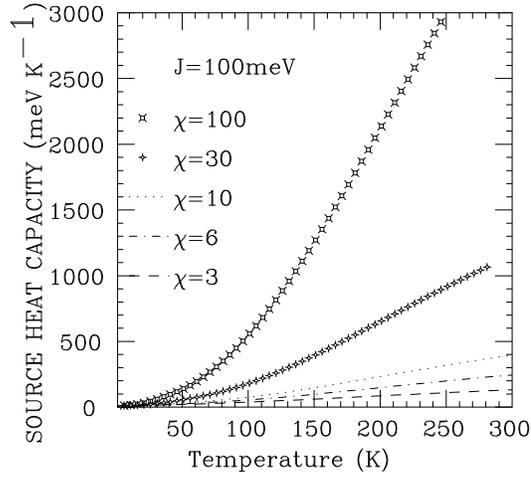}
\caption{Source Term contributions to the heat capacity for the
same parameters as in Fig.4.}
\label{fig:5}       
\end{figure}
\begin{figure}
\centering
\includegraphics[height=10cm,angle=90]{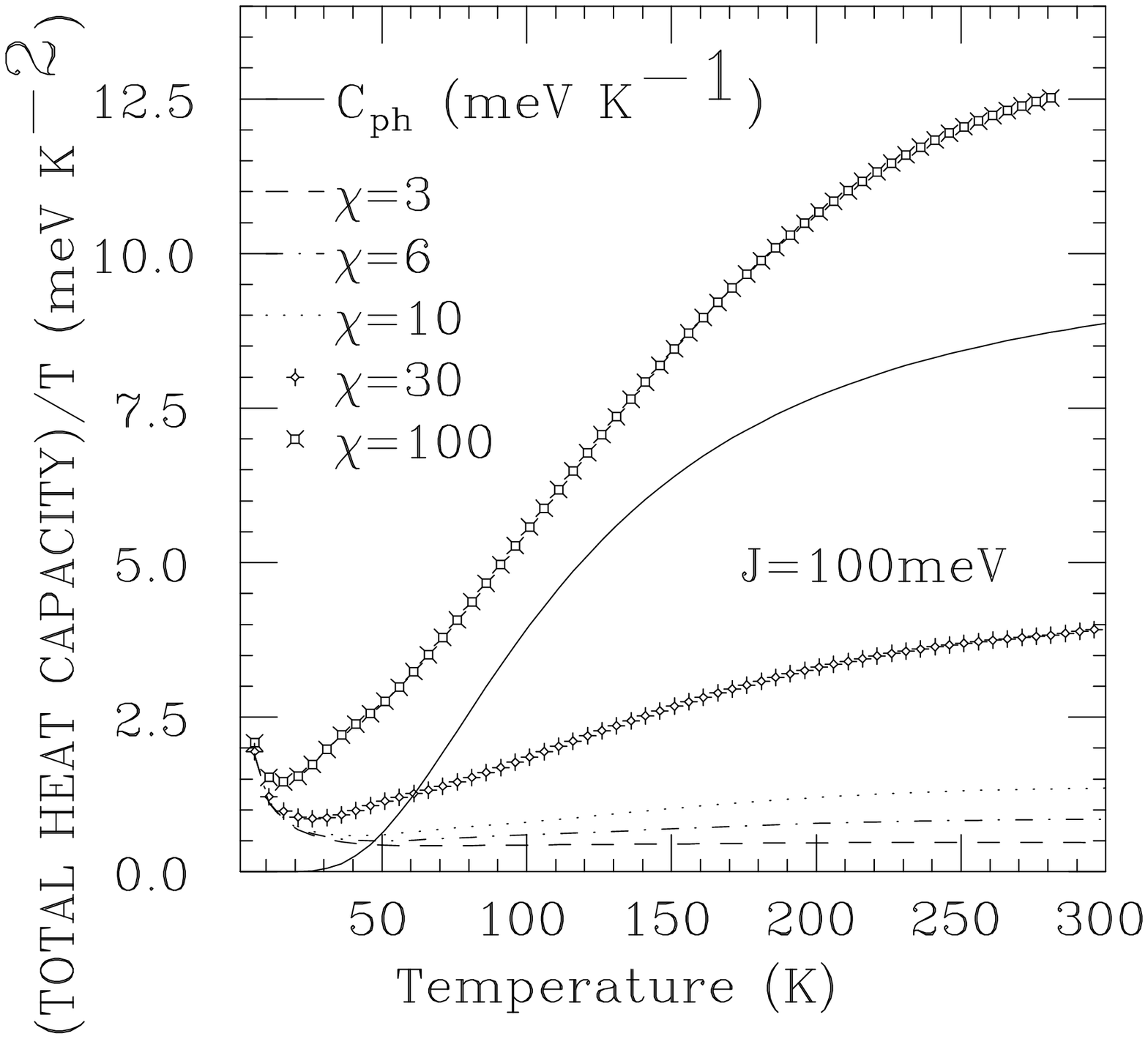}
\caption{Total heat capacity over temperature for the same
parameters as in Fig.4. The phonon heat capacity is also plotted.}
\label{fig:6}       
\end{figure}
\section{ Electron-Phonon Anharmonicity}
\label{sec:5}

So far we have considered a bath of harmonic phonons. Now we face
the following question: what is the effect of the particle-phonon
interaction on the phonon subsystem?

In general, the phonon partition function perturbed by a source
current $j(\tau)$ can be expanded in anharmonic series as:

\begin{eqnarray}
Z_{ph}[j(\tau)] \simeq \, Z_{ph} \biggl(1 + \sum_{l=1}^k (-1)^l
<C^l>_{j(\tau)} \biggr) \quad, \, \nonumber \\ \label{eq:15}
\end{eqnarray}

where the cumulant terms  $<C^l>_{j(\tau)}$ are expectation values
of powers of correlation functions of the perturbing potential.
The averages are meant over the ensemble of the harmonic
oscillators whose partition function is $Z_{ph}$.

\subsection{The Holstein-type Current}
\label{sec:5.1)}

First we consider the general problem of an electron path linearly
coupled to a single oscillator with energy $\omega$ and
displacement $u(\tau)$, through the current $j_x(\tau)=\, -\alpha
x(\tau)$. This type of current models the Holstein interaction
\cite{i7}. In this case odd $k$ cumulant terms vanish and the
lowest order even $k$ cumulants can be straightforwardly derived
\cite{i8}. To obtain a closed analytical expression for the
cumulants to any order we approximate the electron path by its
$\tau$ averaged value:
 $<x(\tau)>\equiv\,{1 \over \beta}\int_0^{\beta} d\tau x(\tau) =\,x_0/a$
and expand the oscillator path in $N_F$ Fourier components:

\begin{eqnarray}
& &u(\tau)=\,u_o + \sum_{n=1}^{N_F} 2\Bigl(\Re u_n \cos( \omega_n
\tau) - \Im x_n \sin( \omega_n \tau) \Bigr)\, \nonumber \\ &
&\omega_n=\,2\pi n/\beta \quad. \label{eq:16}
\end{eqnarray}

Next we choose the  measure of integration

\begin{eqnarray}
\oint Du(\tau)\equiv  & &\biggl({1 \over 2}\biggr)^{2N_F} {{\Bigl(
2\pi \cdot \cdot 2N_F \pi \Bigr)^2} \over {\sqrt{2}
\lambda_M^{(2N_F+1)} }} \int_{-\infty}^{\infty}{du_o} \, \nonumber
\\  \times & &\prod_{n=1}^{N_F} \int_{-\infty}^{\infty} d\Re u_n
\int_{-\infty}^{\infty} d\Im u_n \quad, \label{eq:17}
\end{eqnarray}

being $\lambda_M=\,\sqrt{\pi \hbar^2 \beta/M}$. Such a measure
normalizes the kinetic term in the oscillator field action

\begin{equation}
\oint Du(\tau) exp\Bigl[-{M \over 2}\int_0^\beta d\tau
\dot{u}^2(\tau)\Bigr]\equiv 1 \quad. \label{eq:18}
\end{equation}

Then, using Eqs.~(\ref{eq:15}) -~(\ref{eq:17}), we obtain for the
$k-th$ cumulant

\begin{eqnarray}
& &<C^k>_{N_F}=\,Z^{-1}_{ph} {{(\alpha_{R} \beta \lambda_M)^k (k -
1)!!} \over {k! \pi^{k/2} (\omega \beta)^{k+1}}} \prod_{n=1}^{N_F}
{{(2n\pi)^2} \over {(2n\pi)^2 + (\omega \beta)^2}}\, \nonumber \\
& &\alpha_R=\,\alpha x_0/a  \quad. \label{eq:19}
\end{eqnarray}

Let's set $x_0/a=\,0.1$ in the following calculations thus
reducing the effective coupling $\alpha_R$  by one order of
magnitude with respect to the bare value. However, the trend shown
by the results hereafter presented does not depend on this choice
since $x_0/a$ and $\alpha$ can be varied independently. As the
cumulants should be stable against the number of Fourier
components in the oscillator path expansion, using
Eq.~(\ref{eq:19}) we set the minimum $N_F$ through the condition
$2N_F \pi \gg \omega \beta$. The thermodynamics of the anharmonic
oscillator can be computed by the cumulant corrections to the
harmonic phonon free energy:

\begin{equation}
F^{(k)}(T)=\,-{1 \over \beta} \ln \Bigl[1 +
\sum_{l=1}^{k}<C^{2l}>_{N_F} \Bigr] \quad. \label{eq:20}
\end{equation}

To proceed one needs a criterion to find the temperature dependent
cutoff $k^*$ in the cumulant series.  We feel that, in the low $T$
limit, the third law of thermodynamics may offer the suitable
constraint to determine $k^*$. Then, given $\alpha$ and $\omega$,
the program searches for the cumulant order such that the phonon
heat capacity and the entropy tend to zero in the zero temperature
limit. At any finite temperature T, the constant volume heat
capacity is computed as

\begin{eqnarray}
C_V^{(k)}(T)=\,& &- \Bigl[F^{(k)}(T + 2\Delta) - 2F^{(k)}(T +
\Delta) + F^{(k)}(T) \Bigr] \nonumber \\  \times & &\Bigl({1 \over
\Delta} + {T \over \Delta^2} \Bigr) \quad, \label{eq:21}
\end{eqnarray}

$\Delta$ being the incremental step and $k^*$ is determined as the
minimum value for which the heat capacity converges with an
accuracy of $10^{-4}$. Figs.7(a) and 7(b) show phonon heat
capacity and free energy respectively in the case of a low energy
oscillator for an intermediate value of {\it e-ph} coupling.
Harmonic functions, anharmonic functions with second order
cumulant and anharmonic functions with $k^*$ corrections are
reported on in each figure. The second order cumulant is clearly
inadequate to account for the low temperature trend yielding a
negative phonon heat capacity below $\sim 40K$ while at high $T$
the second order cumulant contribution tends to vanish. Instead,
the inclusion of $k^*$ terms in Eqs.~(\ref{eq:20}),~(\ref{eq:21})
leads to the correct zero temperature limit although there is no
visible anharmonic effect on the phonon heat capacity throughout
the whole temperature range being $C_V^{(k^*)}$ perfectly
superimposed on the harmonic $C_V^h$. Note in Fig.7(b) that the
$k^*$ corrections simply shift downwards the free energy without
changing its slope versus temperature. By increasing $\alpha_R$,
the low T range with wrong (negative) $C_V^{(2)}$ broadens whereas
the $k^*$ contributions permit to fulfill the zero temperature
constraint and substantially lower the phonon free energy. Thus,
for the particular choice of constant (in $\tau$) source current
we find that the {\it e-ph} anharmonicity renormalizes the phonon
partition function although no change occurs in the
thermodynamical behavior of the free energy derivatives.
Anharmonicity is essential to stabilize the system but it leaves
no trace in the heat capacity \cite{gurevich}. Figure 8(a)
displays the $k^*$ temperature dependence for three choices of
{\it e-ph} coupling in the case of a low energy oscillator: while,
at high $T$, the number of required cumulants ranges between six
and ten according to the coupling, $k^*$ strongly grows at low
temperatures reaching the value 100 at $T=\,1K$ for
$\alpha_R=\,60meV \AA^{-1}$. The $k^*$ versus $\alpha_R$ behavior
is depicted in Fig.8(b) for three selected temperatures: at low
$T$ the cutoff strongly varies with the strength of the coupling
while, by enhancing $T$, the number of cumulant terms in the
series is smaller and becomes much less dependent on $\alpha_R$.

\subsection{The SSH-type Current}
\label{sec:5.2)}

Next we turn to the computation of the equilibrium thermodynamics
of the phonon subsystem perturbed by the source current of the
semiclassical SSH model described in Sections 2 and 3. Assuming
that the electron particle path interacts with each of the ${\bar
N}$ oscillators through the coupling $\alpha$ (taken independent
of $i$), we write the $k^{th}$ cumulant term as

\begin{eqnarray}
<C^k>_{j(\tau)}=\, & &Z_{ph}^{-1} \prod_{i=1}^{\bar N} \oint
Du_i(\tau) {1 \over {k!}} \prod_{l=1}^k \Biggl[ \int_0^{\beta}
d\tau_l u_i(\tau_l) j(\tau_l) \Biggr]^{l} \, \nonumber \\ \times &
& exp\Biggl[- \int_0^{\beta} d\tau \sum_{i=1}^{\bar N} {M_i \over
2} \bigl( \dot{u_i}^2(\tau) + \omega_i^2 u_i^2(\tau) \bigr)
\Biggr] \quad, \, \nonumber \\ \label{eq:22}
\end{eqnarray}

where $j(\tau)$ is given in Eq.~(\ref{eq:5}). We take here the 1D
system. Since the oscillators are in fact decoupled in our model
(and anharmonic effects mediated by the electron particle path are
neglected) the behavior of the cumulant terms $<C^k>_{j(\tau)}$
can be studied by selecting a single oscillator having energy
$\omega$ and displacement $u(\tau)$.

As the electron propagator depends on the bare hopping integral we
set as before $J=100meV$. Any electron path yields in principle a
different cumulant contribution. Numerical investigation shows
however that convergent $k-$ order cumulants are achieved by
taking $N_p=\,2$ Fourier components in the electron path expansion
and summing over $\sim 5^{2N_p+1}$ electron paths.

As in the case of {\bf Sect. 5.1}, we truncate the cumulant series
by invoking the third law of thermodynamics to determine the
cutoff $k^*$ in the low temperature limit and by searching for
numerical convergence on the first and second free energy
derivatives at any finite temperature. Again, we can start our
analysis from Eq.~(\ref{eq:20}) after checking that odd $k$
cumulants yield vanishing contributions. Now however the picture
of the anharmonic effects changes drastically. The {\it e-ph}
coupling strongly modifies the shape of the heat capacity and free
energy plots with respect to the harmonic result as it is seen in
Figs.9(a) and 9(b) respectively. The heat capacity versus
temperature curves show a peculiar peak above a threshold value
$\alpha \sim 10meV \AA^{-1}$ which clearly varies according to the
energy of the harmonic oscillator. Here we set $\omega=\,20meV$ to
emphasize the size of the anharmonic effects on a low energy
oscillator. By enhancing $\alpha$ the height of the peak grows and
the bulk of the anharmonic effects on the heat capacity is shifted
towards lower $T$. At $\alpha \sim 60meV \AA^{-1}$ the crossover
temperature is around 100K. Note that the size of the anharmonic
enhancement is $\sim 10$ times the value of the harmonic
oscillator heat capacity at $T=\,100K$. It is worth noting that
previous numerical studies of a classical one dimensional
anharmonic model undergoing a Peierls instability \cite{allen1}
also found a specific heat peak as a signature of anharmonicity.
However such a large anharmonic effect on the phonon subsystem is
partly covered in the total heat capacity by the source action
$A(j(\tau))$ and mainly by the hopping potential $V(x(\tau))$
contributions analysed in the previous Section.

Taken for instance a bath of ten low energy oscillators with
$\omega=\,20meV$, setting $\alpha \sim 60meV \AA^{-1}$ which
implies an effective coupling $\chi \sim 700meV^3$ (last of
Eqs.~(\ref{eq:5})) we get a source heat capacity a factor two
larger than the harmonic phonon heat capacity at temperatures of
order $100K$. Thus the anharmonic peak, although substantially
smeared by the electronic contributions to the total heat
capacity, should still appear in systems with low energy phonons
and sizeable {\it e-ph} coupling to which the SSH Hamiltonian
applies. Let's focus on this point.

The total heat capacity is given by the phonon contribution plus a
{\it source} heat capacity which includes both the electronic
contribution (related to the electron hopping integral) and the
contribution due to the source action (the latter being $\propto
\alpha^2$). Fig.10(a) shows the comparison between the anharmonic
phonon heat capacity ($C_V^{anh}$) and the source heat capacity,
here termed $C_V^{e-p}$ to emphasize the dependence both on the
electronic and on the {\it e-ph} coupling terms. $C_V^{e-p}$ is
computed as described above setting $\alpha= 21.74 meV \AA^{-1}$
\cite{i4}. Also the total heat capacity ($C_V^{tot}= C_V^{anh} +
C_V^{e-p}$) is shown in Fig.10(a). At low temperatures,
$C_V^{e-p}$ yields the largest effect mainly due to the electronic
hopping while at high $T$, {} $C_V^{e-p}$ prevails as the source
action becomes dominant. In the intermediate  range ($T \in
[90,210]$) the anharmonic phonons provide the highest contribution
although their characteristic peak is substantially smeared in the
total heat capacity by the source term background. Fig.10(b)
compares $C_V^{anh}$ and $C_V^{e-p}$ for two increasing values of
$\alpha$: while the anharmonic peak shifts downwards (along the
$T$ axis) by enhancing $\alpha$, $C_V^{anh}$ remains larger than
$C_V^{e-p}$ in a temperature range which progressively shrinks due
to the strong dependence of the source action on the strength of
the {\it e-ph} coupling. Finally, we observe that the low
temperature upturn displayed in the {\it total heat capacity over
{} T} ratio discussed above is not affected by the inclusion of
phonon anharmonic effects which tend to become negligible at low
temperatures.

\section{ Conclusions}
\label{sec:6}

Mapping the real space Su-Schrieffer-Heeger model onto the time
scale I have developed a semiclassical version of the interacting
model Hamiltonian in one and two dimensions suitable to be
attacked by path integrals methods. The acoustical phonons of the
standard SSH model have been replaced by a set of oscillators
providing a bath for the electron interacting with the
displacements field. Time retarded interactions are naturally
introduced in the formalism through the source action $A(x(\tau))$
which depends quadratically on the bare {\it e-ph} coupling
strength $\alpha$. Via calculation of the electronic motion path
integral, the partition function can be derived in principle for
any value of $\alpha$ thus avoiding those limitations on the {\it
e-ph} coupling range which burden any perturbative method.
Particular attention has been paid to establish a reliable and
general procedure which allows one to determine those input
parameters intrinsic to the path integral formalism. It turns out
that a large number of paths is required to carry out low
temperature calculations which therefore become highly time
consuming. The physical parameters have been specified to a narrow
band system and the behavior of some thermodynamical properties,
free energy and heat capacity, has been analysed for some values
of the effective coupling strength lying both in the weak and in
the strong coupling regime. We find, both in 1D and 2D, a peculiar
upturn in the low temperature plots of the heat capacity over
temperature ratio indicating that a glass-like behavior can arise
in the linear chain as a consequence of a time dependent
electronic hopping with variable range.

According to our integration method (Eq.~(\ref{eq:7})), at any
temperature, a specific set of Fourier coefficients defines the
ensamble of relevant particle paths over which the hopping
potential $V\bigl(x(\tau)\bigr)$ is evaluated. This ensamble is
therefore $T$ dependent. However, given a single set of path
parameters one can monitor the $V\bigl(x(\tau)\bigr)$ behavior
versus $T$. I find that the hopping decreases (as expected) by
lowering $T$ but its value remains appreciable also at low
temperatures ($\preceq 20K$ in 2D and $\preceq 10K$ in 1D). Since
the $d\tau$ integration range is larger at lower temperatures, the
overall hopping potential contribution to the total action is
relevant also at low $T$. It is precisely this property which is
responsible for the anomalous upturn in the heat capacity linear
coefficient. Further investigation also reveals that the upturn
persists both in the extremely narrow ($J \sim 10meV$) and in the
wide band ($J \sim 1eV$) regimes. Moreover, the upturn is not
modified by the inclusion of electron-phonon anharmonicity in the
phonon subsystem.

The presented computational method accounts for a variable range
hopping on the $\tau$ scale which corresponds physically to
introduce some degree of disorder along the linear chain. This
feature makes my model more general than the standard SSH
Hamiltonian (Eq.~(\ref{eq:1})) with only real space nearest
neighbor hops. While hopping type mechanisms have been suggested
\cite{kivel} to explain the striking conducting properties of
doped polyacetylene at low temperatures I am not aware of any
other direct calculation of the specific heat in the SSH model.
Since the latter quantity directly probes the density of states
and integrating over $T$ the {\it specific heat over T} ratio one
can have access to the experimental entropy, this method may
provide a new approach to analyse the transition to a disordered
state which indeed exists in polymers. In this connection it is
also worth noting that the low $T$ upturn in the specific heat
over $T$ ratio is a peculiar property of glasses
\cite{zeller,anderson} in which tunneling states for atoms (or
group of atoms) provide a non magnetic internal degree of freedom
in the potential structure \cite{zawa,io91}.

\begin{figure}
\centering
\includegraphics[height=10cm,angle=90]{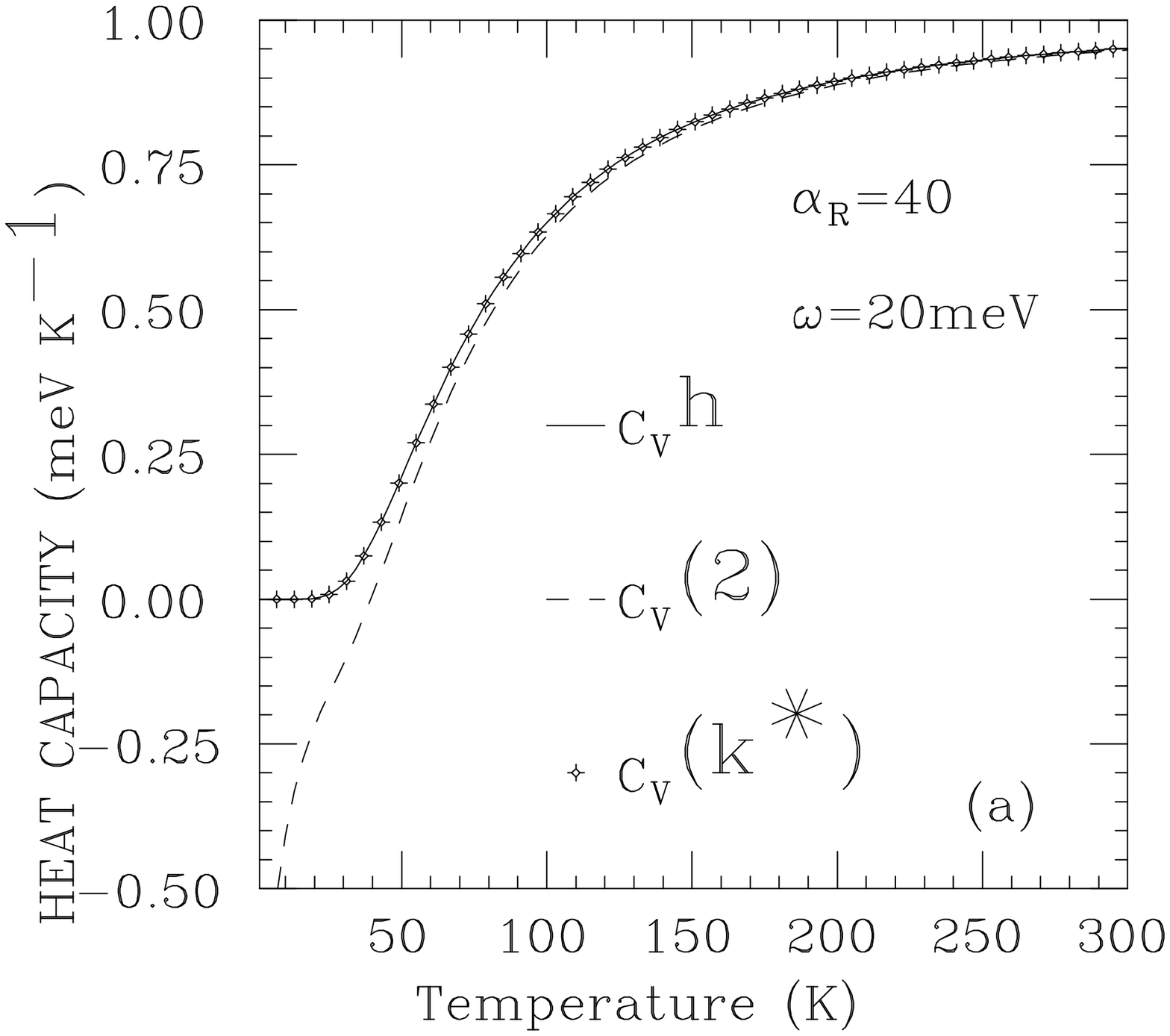}
\includegraphics[height=10cm,angle=90]{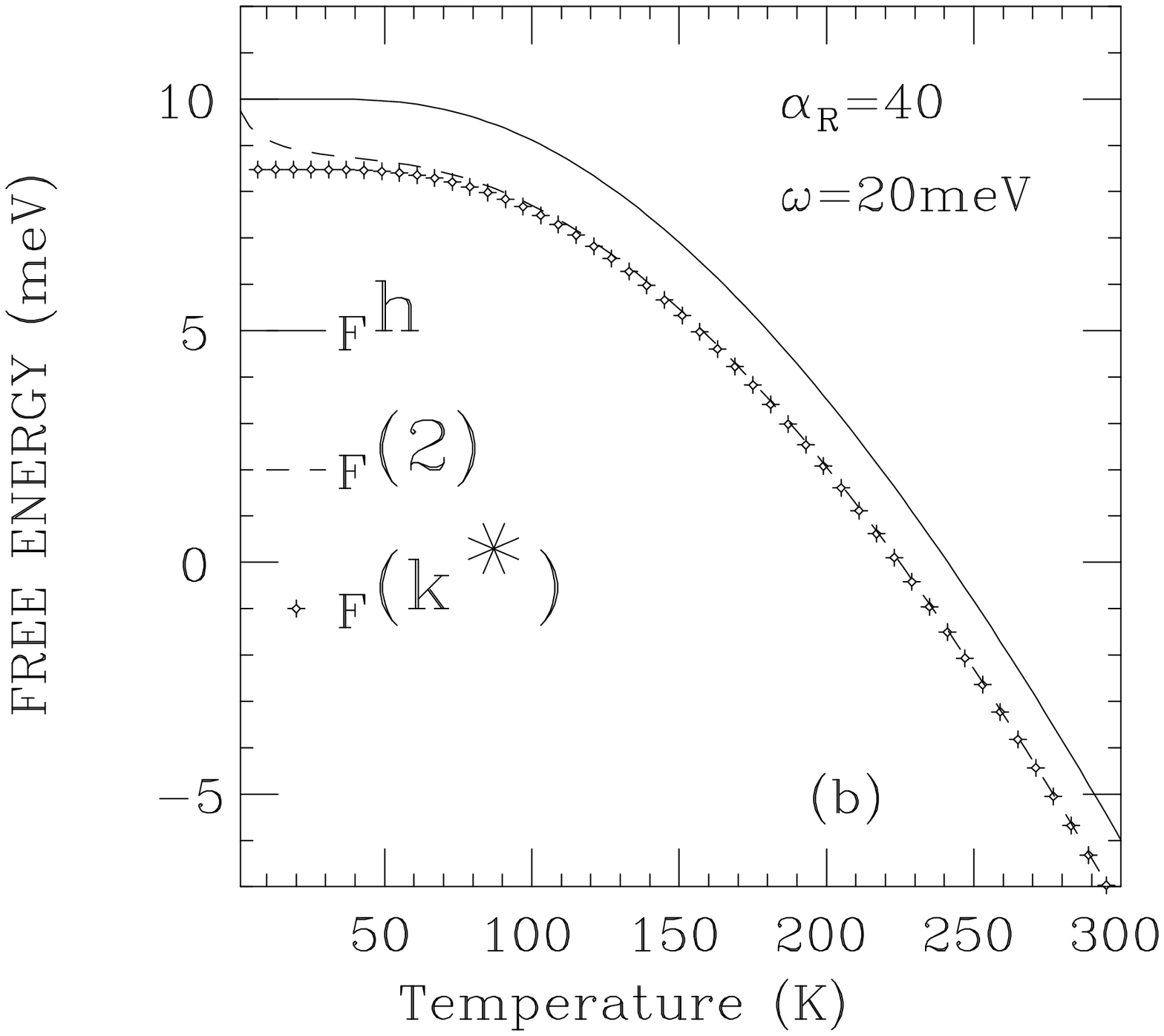}
\caption{(a) 1D Phonon heat capacity and (b) 1D Phonon free energy
calculated in i) the harmonic model, ii) anharmonic model with
second order cumulant, iii) anharmonic model with $k^*$ cumulants
(see text).  $\alpha_R$ is the effective {\it e-ph} coupling in
units $meV \AA^{-1}$ and $\omega$ is the phonon energy.}
\label{fig:7}       
\end{figure}
\begin{figure}
\centering
\includegraphics[height=10cm,angle=90]{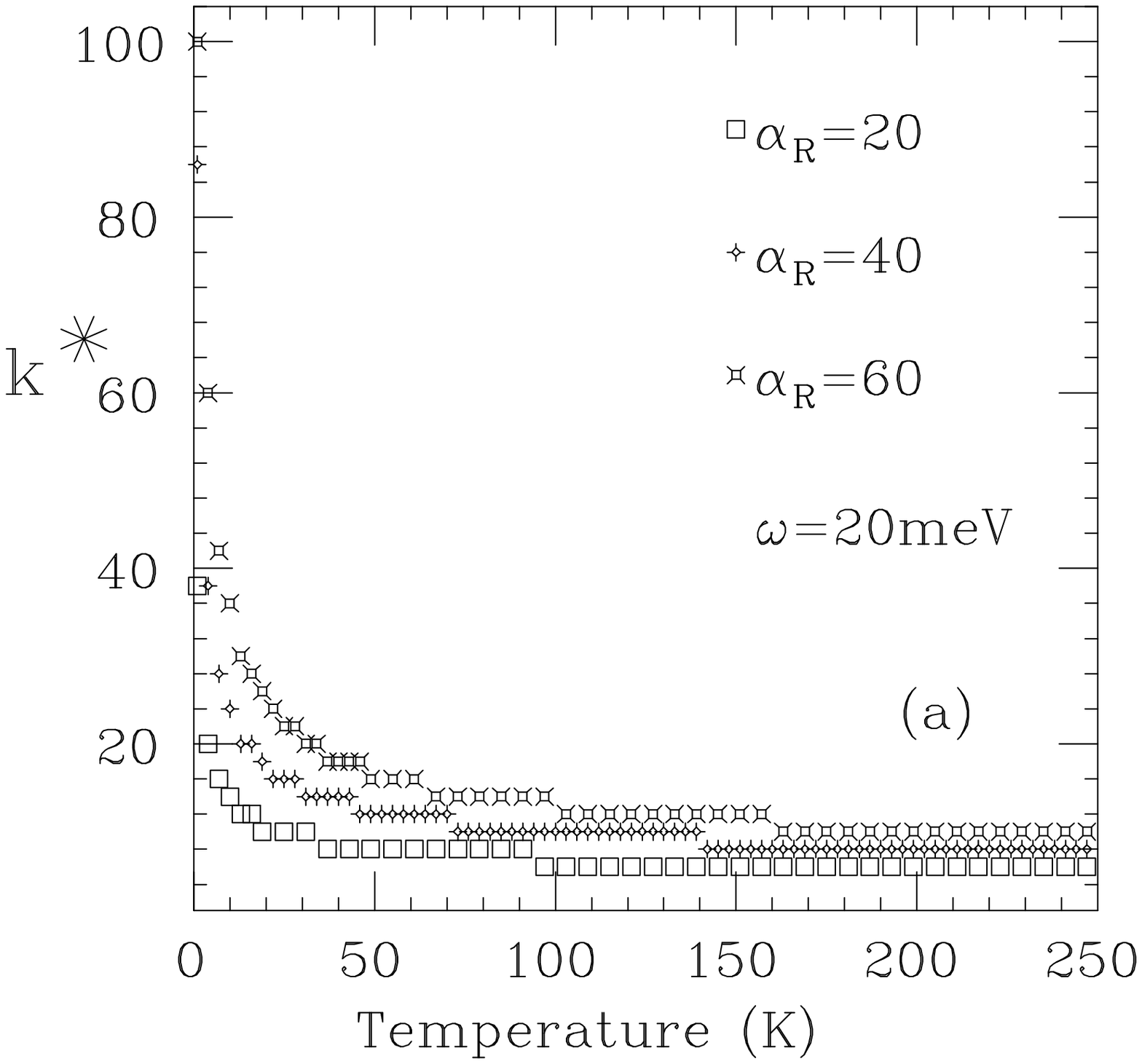}
\includegraphics[height=10cm,angle=90]{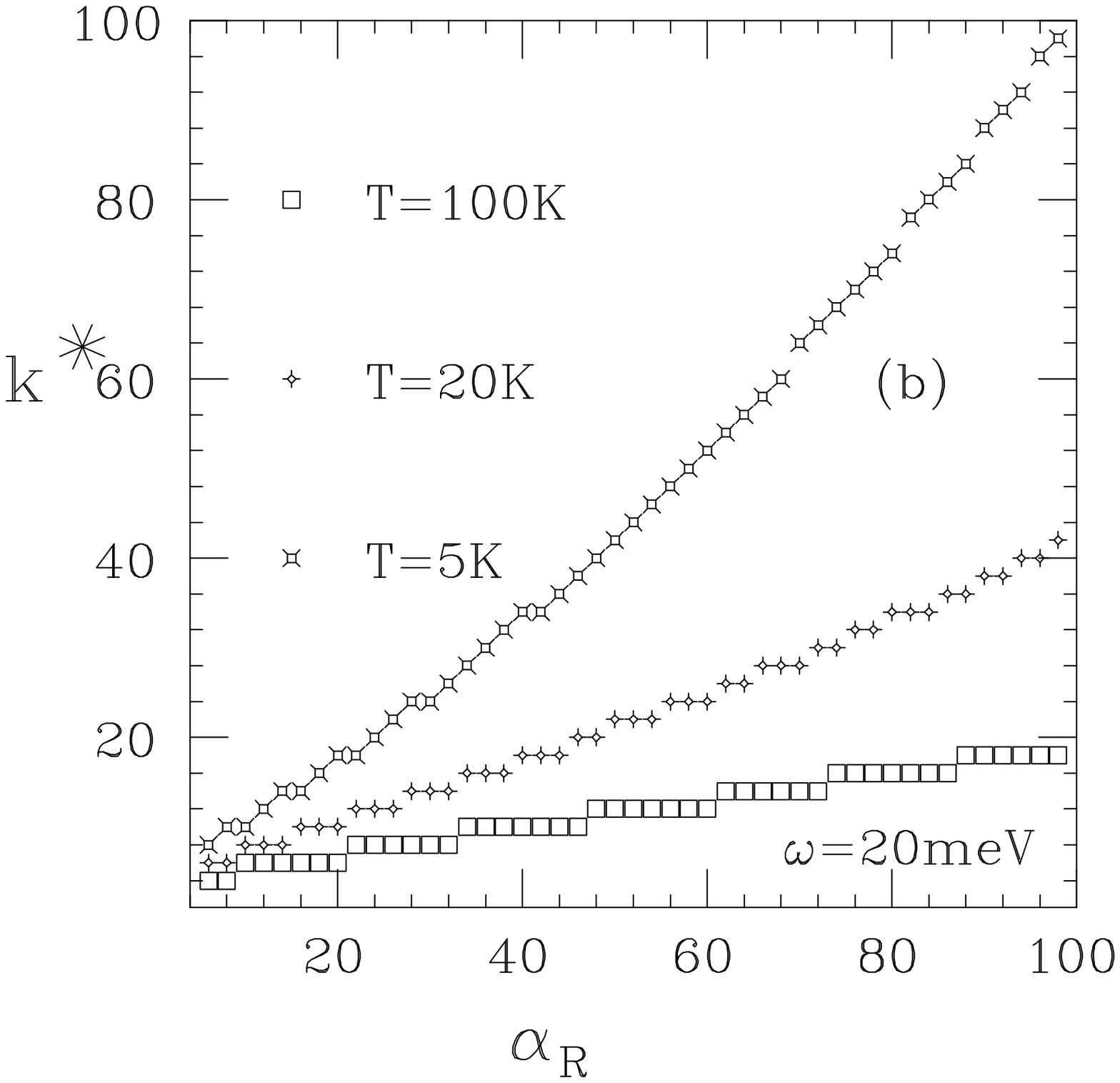}
\caption{(a) Number of cumulants required to obtain a convergent
phonon heat capacity at any temperature for different choices of
{\it e-ph} couplings. (b) Number of cumulants yielding a
convergent phonon heat capacity at any {\it e-ph} coupling for
three selected temperatures.}
\label{fig:8}       
\end{figure}
\begin{figure}
\centering
\includegraphics[height=10cm,angle=90]{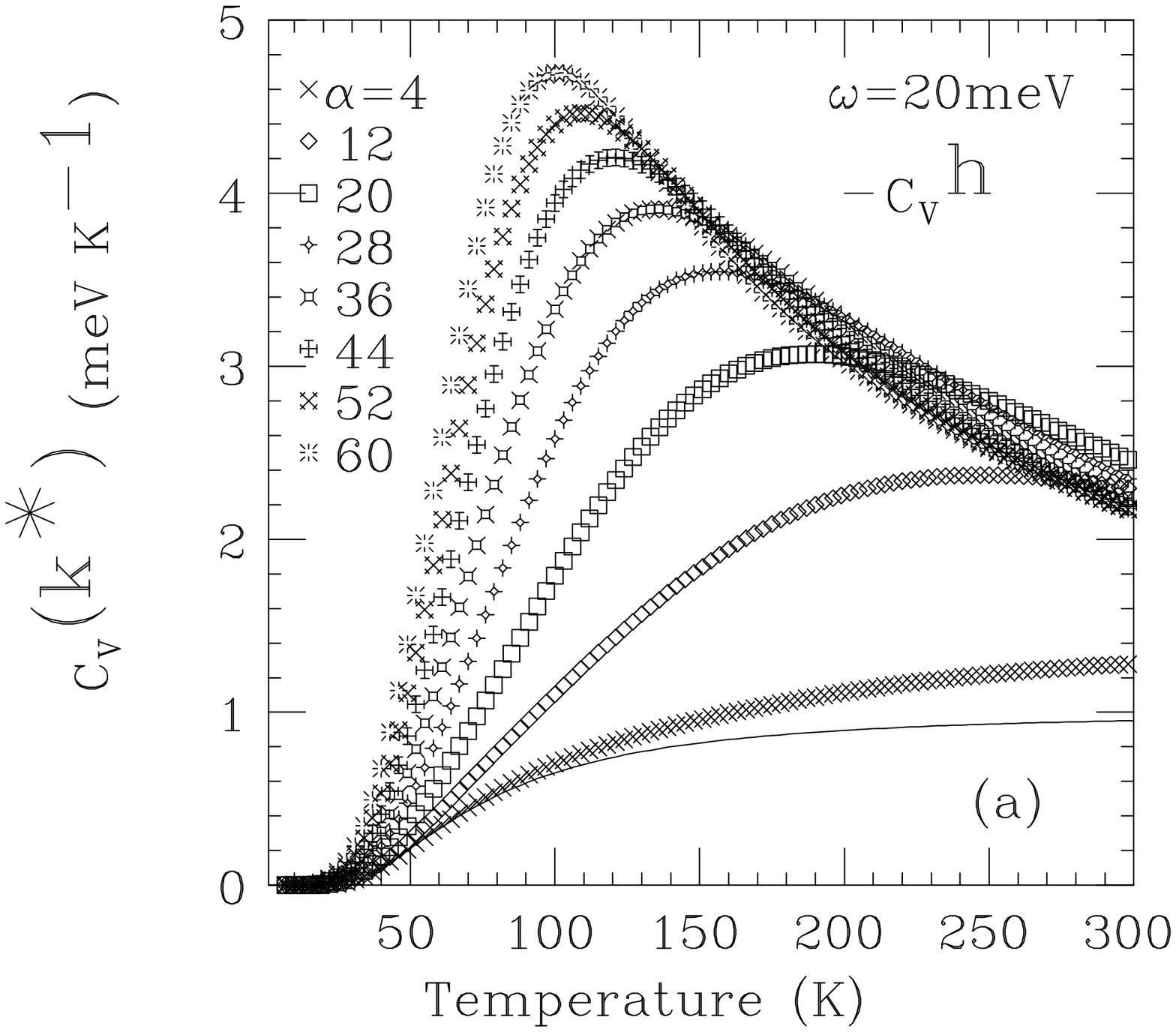}
\includegraphics[height=10cm,angle=90]{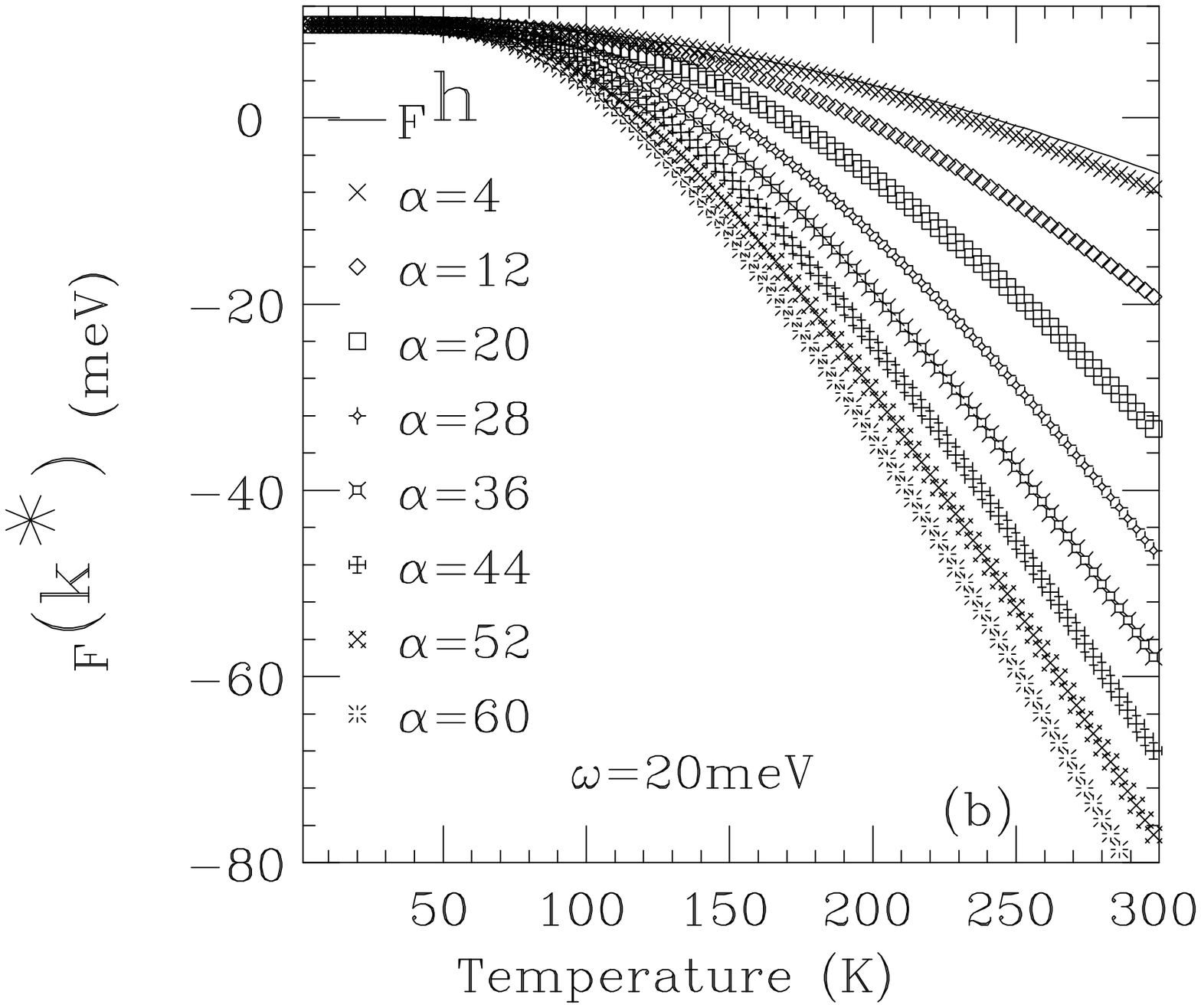}
\caption{1D Anharmonic (a) phonon heat capacity and (b) free
energy versus temperature for eight values of {\it e-ph} coupling.
The harmonic plots are also reported on. A low energy oscillator
is assumed. }
\label{fig:9}       
\end{figure}
\begin{figure}
\centering
\includegraphics[height=10cm,angle=90]{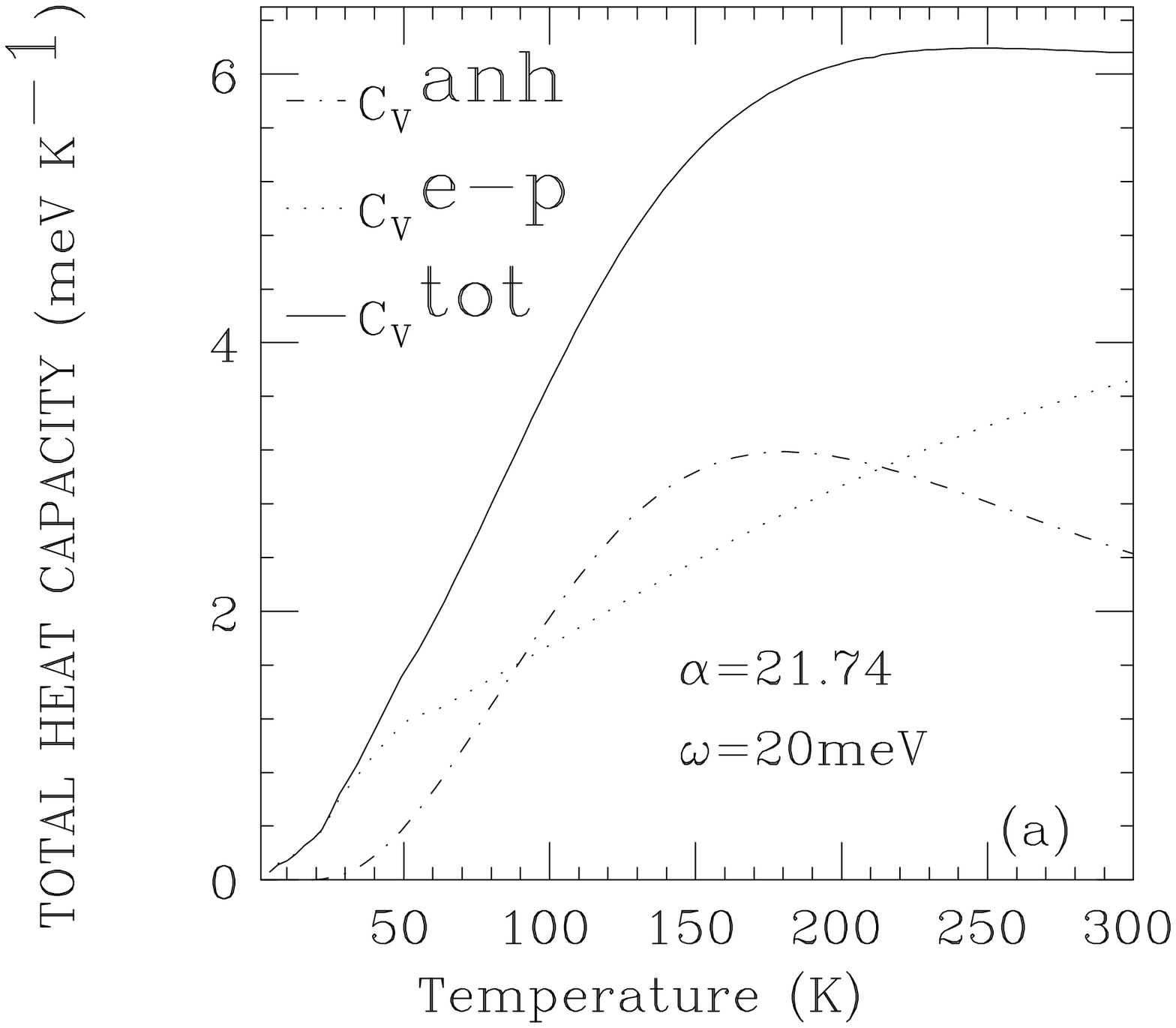}
\includegraphics[height=10cm,angle=90]{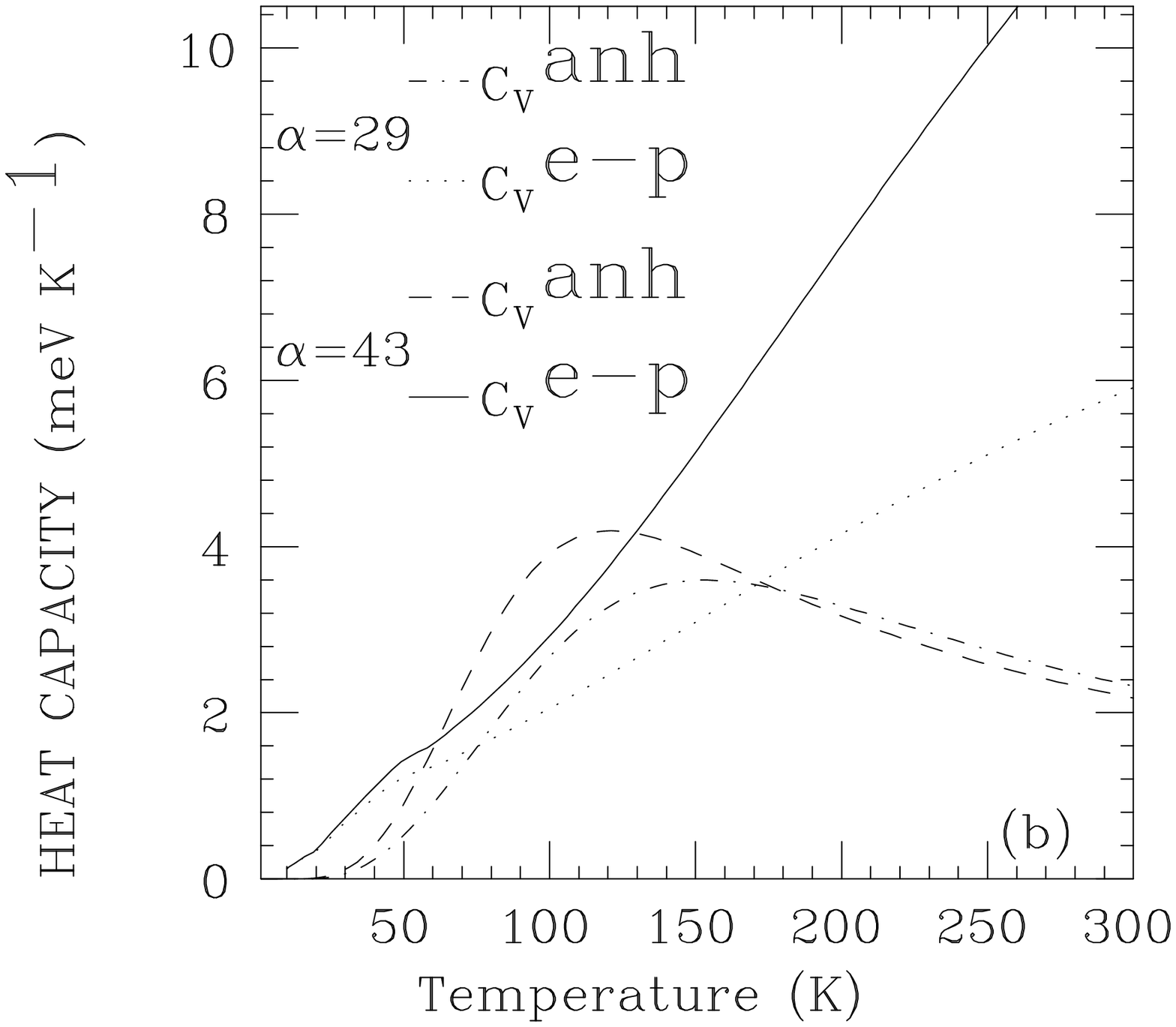}
\caption{(a) Total Heat Capacity versus temperature in the 1D
Su-Schrieffer-Heeger model. The contributions due to anharmonic
phonons ($C_V^{anh}$) and electrons {\it plus} electron-phonon
interactions ($C_V^{e-p}$) are plotted separately. The largest
$\alpha$ of Fig.1 is assumed. (b) $C_V^{anh}$ and $C_V^{e-p}$ for
two values of $\alpha$ in units $meV \AA^{-1}$. $\omega=\,20meV$.
}
\label{fig:10}       
\end{figure}
%

\input{refpol06}

\printindex
\end{document}

%% file: refpol06.tex
%
%

%
%